\documentclass[12pt]{article}
\usepackage{euler}
\usepackage{amsmath}
\usepackage{amssymb}
\usepackage{amstext}
\usepackage{amscd}
\usepackage{amsfonts}

\title{{\bf Modelling and Analysis of Fractional Order Systems using
Ultradistributions }
\thanks{\it{This work was partially supported by Consejo
Nacional
de Investigaciones Cient\'{\i}ficas
Argentina.}}}
\author{C.M.Grunfeld and M.C.Rocca\\
Departamento de F\'{\i}sica, Fac. de Ciencias Exactas,\\
Universidad Nacional de La Plata.\\
C.C. 67 (1900) La Plata. Argentina.}

\date{March 23, 2009}

\begin{document}

\maketitle

\vspace{-5mm}

\begin{abstract}
In this paper we introduce a new mathematical tool to solve  fractional
equations representing models of fractional systems : The Ultradistributions.\\
Ultradistributions permit us to unify the notion of integral and derivative in one only
operation. Several examples of application of the results obtained are given.


\end{abstract}

\newpage

\renewcommand{\theequation}{\arabic{section}.\arabic{equation}}

\section{Introduction}

\setcounter{equation}{0}

The use of fractional calculus for modelling physical systems has been considered in
many works. See for example \cite{tq1,tq2,tq3}. We can find also works
dealing with the application of this mathematical tool in control theory
\cite{tq4,tq5,tq6,tq7}..

Moreover, there are many physical systems that can be described by 
means of a fractional calculus. Some examples are: chaos \cite{tq8},
long electric lines \cite{tq9}, electrochemical process \cite{tq10} and 
dielectric polarization \cite{tq11}.

In this paper we want to introduce a new mathematical framework to solve fractional
equations representing models of fractional systems which was not treated in none of the
previous works: The Ultradistributions.

The paper is organized as follow: in section 2  we introduce definition
of fractional derivation and integration. In section 3 
we give some examples of application
of the formulae of section 2 using 
the Fourier Transform and 
the one-side Laplace Transform. In section 3 we present 
a circuital application. Finally in section 4 we discuss 
the results obtained in sections 1,2 and 3.

\section{Fractional Calculus}

\setcounter{equation}{0}
The purpose of this sections is to introduce definition of fractional
derivation and integration given in ref. \cite{tp1}.
This definition unifies the notion of integral and derivative in one only
operation.
Let $\hat{f}(x)$ a distribution of exponential type and $F(\Omega)$
the complex Fourier transformed Tempered Ultradistribution.
Then:
\begin{equation}
\label{ep2.1}
F(\Omega)=U[\Im(\Omega)]\int\limits_0^{\infty}\hat{f}(x) e^{j\Omega x}\;dx-
U[-\Im(\Omega)]\int\limits_{-\infty}^0\hat{f}(x) e^{j\Omega x}\;dx
\end{equation}
($U(x)$ is the Heaviside step function)
and
\begin{equation}
\label{ep2.2}
\hat{f}(x)=\frac {1} {2\pi}\oint\limits_{\Gamma}F(\Omega)
e^{-j\Omega x}\;d\Omega
\end{equation}
where the contour $\Gamma$ surround all singularities of
$F(\Omega)$ and runs parallel to real axis from $-\infty$ to
$\infty$ above the real axis and from $\infty$ to $-\infty$
below the real axis.
According to \cite{tp1} the fractional derivative of $\hat{f}(x)$ is given by
\begin{equation}
\label{ep2.3}
\frac {d^{\lambda}\hat{f}(x)} {dx^{\lambda}}=\frac {1} {2\pi}\oint\limits_{\Gamma}
(-j\Omega)^{\lambda}
F(\Omega)
e^{-j\Omega x}\;d\Omega+
\oint\limits_{\Gamma}(-j\Omega)^{\lambda}a(\Omega)
e^{-j\Omega x}\;d\Omega
\end{equation}
Where $a(\Omega)$ is entire analytic and rapidly decreasing.
If $\lambda=-1$, $d^{\lambda}/dx^{\lambda}$ is the inverse of the derivative
(an integration). In this case the second term of the right side of (\ref{ep2.3})
gives a primitive of $\hat{f}(x)$. Using Cauchy's theorem the additional
term is 
\begin{equation}
\label{ep2.4}
\oint \frac {a(\Omega)} {\Omega}e^{-j\Omega x} d\Omega=
2\pi a(0)
\end{equation}
Of course, an integration should give a primitive plus an arbitrary constant.
Analogously when $\lambda=-2$ (a double iterated integration) we have
\begin{equation}
\label{ep2.5}
\oint \frac {a(\Omega)} {{\Omega}^2}e^{-j\Omega x} d\Omega=
\gamma+\delta x
\end{equation}
where $\gamma$ and $\delta$ are arbitrary constants.
With the change of variables $s=-j\Omega$ formulae 
(\ref{ep2.1}) and (\ref{ep2.2}) can be writen as:
\begin{equation}
\label{ep2.6}
G(s)=U[\Re(s)]\int\limits_0^{\infty}\hat{f}(x) e^{-sx}\;dx-
U[-\Re(s)]\int\limits_{-\infty}^0\hat{f}(x) e^{-sx}\;dx
\end{equation}
and
\begin{equation}
\label{ep2.7}
\hat{f}(x)=\frac {1} {2\pi i}\oint_{\Gamma} G(s) e^{sx}\;ds
\end{equation}
where the contour $\Gamma$ surround all singularities of
$G(S)$ and runs parallel to imaginary axis from $-j\infty$ to
$j\infty$ to the right of the imaginary axis and from $j\infty$ to $-j\infty$
to the left of  the imaginary axis. Formula (\ref{ep2.6}) represents 
the two-sided Lapnace Transform. The fractional derivative is now:
\begin{equation}
\label{ep2.8}
\frac {d^{\lambda}\hat{f}(x)} {dx^{\lambda}}=\frac {1} {2\pi i}\oint\limits_{\Gamma}
s^{\lambda}G(s) e^{s x}\;ds +
\oint\limits_{\Gamma}s^{\lambda}a(s)
e^{sx}\;ds
\end{equation}
For the one-side Laplace Transform we have
\begin{equation}
\label{ep2.9}
G(s)=U[\Re(s)]\int\limits_0^{\infty}\hat{f}(x) e^{-sx}\;dx
\end{equation}
\begin{equation}
\label{ep2.10}
\hat{f}(x)=\frac {1} {2\pi j} \int\limits_{a-j\infty}^{a+j\infty}
G(s) e^{sx}\;ds 
\end{equation}
and for the fractional derivative:
\begin{equation}
\label{ep2.11}
\frac {d^{\lambda}\hat{f}(x)} {dx^{\lambda}}=
\frac {1} {2\pi j} \int\limits_{a-j\infty}^{a+j\infty}
s^{\lambda}
G(s) e^{sx}\;ds 
\end{equation}

\section{Examples}

\setcounter{equation}{0}

In this section we give some examples of the application
of formulae of the precedent section. At first using 
the Fourier Transform and at second place using
the one-side Laplace Transform.

\subsection*{The Fourier Transform}

Let $U(x)$ be the Heaviside step function.
\begin{equation}
\label{ep3.1}
\hat{f}(x)=U(x)\;\;\;;\;\;\;
F(\Omega)=U[\Im(\Omega)]\int\limits_0^{\infty}e^{-j\Omega x}\;dx=
\frac {jU[\Im(\Omega)]} {\Omega}
\end{equation}
The fractional derivative is:
\[\frac {d^{\lambda}U(x)} {dx^{\lambda}}=
\frac {je^{-\frac {j\pi\lambda} {2}}} {2\pi}\oint\limits_{\Gamma}
U[\Im(\Omega)]{\Omega}^{\lambda-1}e^{-j\Omega x}\;d\Omega+
\oint\limits_{\Gamma} {\Omega}^{\lambda} 
a(\Omega)e^{-j\Omega x}\;d\Omega=\]
\begin{equation}
\label{ep3.2}
\frac {je^{\frac {-j\pi\lambda} {2}}} {2\pi}
\int\limits_{-\infty}^{\infty} (\omega+j0)^{\lambda-1}
e^{-j\omega x}\;d\omega+
\oint\limits_{\Gamma} {\Omega}^{\lambda} 
a(\Omega)e^{-j\Omega x}\;d\Omega
\end{equation}
With the use of the result (see ref.\cite{tp7})
\begin{equation}
\label{ep3.3}
\int\limits_{-\infty}^{\infty} (\omega+j0)^{\lambda-1}
e^{-j\omega x}\;d\omega=-2\pi j\frac {e^{\frac {i\pi\lambda} {2}}}
{\Gamma(1-\lambda)} x_+^{-\lambda}
\end{equation}
we obtain:
\begin{equation}
\label{ep3.4}
\frac {d^{\lambda}U(x)} {dx^{\lambda}}=
\frac {x_+^{-\lambda}} {\Gamma(1-\lambda)}+
\oint\limits_{\Gamma} {\Omega}^{\lambda} 
a(\Omega)e^{-j\Omega x}\;d\Omega
\end{equation}
When $\lambda=n$ 
\begin{equation}
\label{ep3.5}
\left.\frac {x_+^{-\lambda}} {\Gamma(1-\lambda)}\right|_{\lambda=n}=
\delta^{(n-1)}(x)
\end{equation}
\begin{equation}
\label{ep3.6}
\oint\limits_{\Gamma} {\Omega}^n
a(\Omega)e^{-j\Omega x}\;d\Omega=0
\end{equation}
and we have the ordinary derivative:
\begin{equation}
\label{ep3.7}
\frac {d^n U(x)} {dx^n}=\delta^{(n-1)}(x)
\end{equation}
When $\lambda=-n$ 
\begin{equation}
\label{ep3.8}
\frac {d^{-n}U(x)} {dx^{-n}}=\frac {x_+^n} {n!}+a_0+a_1x+
a_2x^2+\cdot\cdot\cdot+a_{n-1}x^{n-1}
\end{equation}
which is a n-times iterated integral.

Let $\delta(x)$ the Dirac's delta distribution.
For it we have:
\begin{equation}
\label{ep3.9}
\hat{f}(x)=\delta(x)\;\;\;;\;\;\;
F(\Omega)=\frac {Sgn[\Im(\Omega)]} {2}
\end{equation}
The fractional derivative is:
\begin{equation}
\label{ep3.10}
\frac {d^{\lambda}\delta(x)} {dx^{\lambda}}=
\frac {x_+^{-\lambda-1}} {\Gamma(-\lambda)}+
\oint\limits_{\Gamma} {\Omega}^{\lambda} 
a(\Omega)e^{-j\Omega x}\;d\Omega
\end{equation}
When $\lambda=n$:
\begin{equation}
\label{ep3.11}
\frac {d^n \delta(x)} {dx^n}=\delta^{(n)}(x)
\end{equation}
and when $\lambda=-n$:
\begin{equation}
\label{ep3.12}
\frac {d^{-n}\delta(x)} {dx^{-n}}=\frac {x_+^{n-1}} {(n-1)!}+a_0+a_1x+
a_2x^2+\cdot\cdot\cdot+a_{n-1}x^{n-1}
\end{equation}
Let us consider now the fractional derivative of  $e^{jbx}$
\begin{equation}
\label{ep3.13}
\hat{f}(x)=e^{jbx}\;\;\;;\;\;\;F(\Omega)=\frac {j} {\Omega+b}
\end{equation}
We have:
\begin{equation}
\label{ep3.14}
\frac {d^{\lambda}e^{jbx}} {dx^{\lambda}}=
 \frac {j} {2\pi}\oint\limits_{\Gamma}
\frac {(-j\Omega)^{\lambda}e^{-j\Omega x}} {\Omega+b}\;d\Omega+
\oint\limits_{\Gamma}{\Omega}^{\lambda}a(\Omega)e^{-j\Omega x}\;
d\Omega=
\end{equation}
\[\frac {ie^{\frac {-i\pi\lambda} {2}}} {2\pi}
\int\limits_{-\infty}^{\infty}\frac {(\omega+j0)^{\lambda}}
{\omega+b+j0} e^{-j\omega x}d\omega-
\frac {ie^{\frac {-i\pi\lambda} {2}}} {2\pi}
\int\limits_{-\infty}^{\infty}\frac {(\omega-j0)^{\lambda}}
{\omega+b-j0} e^{-j\omega x}d\omega+\]
\begin{equation}
\label{ep3.15}
\oint\limits_{\Gamma}{\Omega}^{\lambda}a(\Omega)e^{-j\Omega x}\;
d\Omega
\end{equation}
From ref.\cite{tp10} we obtain:
\[\int\limits_{-\infty}^{\infty}\frac {(x+\gamma)^{\lambda}}
{x+\beta}e^{-ipx}dx=\]
\begin{equation}
\label{ep3.16}
2\pi U(p)
\frac {e^{\frac {-j\pi} {2}(1-\lambda)}}
{\Gamma(1-\lambda)}p^{-\lambda}
e^{i\beta p}\phi[-\lambda,1-\lambda,j(\gamma-\beta)p]
\end{equation}
where $\phi$ is the confluent hypergeometric function. 
Thus the fractional derivative is:
\begin{equation}
\label{ep3.17}
\frac {d^{\lambda}e^{jbx}} {dx^{\lambda}}=
\frac {(x+j0)^{-\lambda}} {\Gamma(1-\lambda)}
\phi(1,1-\lambda,jbx)+
\oint\limits_{\Gamma}{\Omega}^{\lambda}a(\Omega)e^{-j\Omega x}\;
d\Omega
\end{equation}
With the use of equality:
\begin{equation}
\label{ep3.18}
\phi(1,1-\lambda,jbx)=(jbx)^{\lambda}e^{jbx}
\left[\Gamma(1-\lambda)+\lambda\Gamma(-\lambda,jbx)\right]
\end{equation}
where $\Gamma(z_1,z_2)$ is the incomplete gamma function,
(\ref{ep3.17}) takes the form:
\[\frac {d^{\lambda}e^{jbx}} {dx^{\lambda}}=
(jb)^{\lambda}e^{jbx}\left[1+\frac {\lambda} {\Gamma(1-\lambda)}
\Gamma(-\lambda,jbx)\right]+\]
\begin{equation}
\label{ep3.19}
\oint\limits_{\Gamma}{\Omega}^{\lambda}a(\Omega)e^{-j\Omega x}\;
d\Omega
\end{equation}
When $\lambda=n$ 
\begin{equation}
\label{ep3.20}
\frac {d^ne^{jbx}} {dx^n}=(jb)^ne^{jbx} 
\end{equation}
and when $\lambda=-n$:
\begin{equation}
\label{ep3.21}
\frac {d^{-n}e^{jbx}} {dx^{-n}}=(jb)^{-n}e^{jbx}+a_0+a_1x+
\cdot\cdot\cdot+a_{n-1}x^{n-1}
\end{equation}

\subsection*{The Laplace Transform}

If we use the one-side Laplace transform to evaluate 
the fractional derivative of $U(x)$,then:
\begin{equation}
\label{ep3.22}
\hat{f}(x)=U(x)\;\;\;;\;\;\;G(s)=U[\Re(s)]
\int\limits_0^{\infty}e^{-sx}dx=
\frac {U[\Re(s)]} {s}
\end{equation}
and as a consequence:
\begin{equation}
\label{ep3.23}
\frac {d^{\lambda}U(x)} {dx^{\lambda}}=
\frac {1} {2\pi j} \int\limits_{a-j\infty}^{a+j\infty}
U[\Re(s)]s^{\lambda-1}e^{sx}\;ds=
\end{equation}
\begin{equation}
\label{ep3.24}
\frac {e^{-ax}} {2\pi}\int\limits_{-\infty}^{\infty}
\frac {e^{jsx}} {(a+js)^{1-\lambda}}ds=
\frac {x_+^{-\lambda}} {\Gamma(1-\lambda)}
\end{equation}
\begin{equation}
\label{ep3.25}
\frac {d^{\lambda}U(x)} {dx^{\lambda}}=
\frac {x_+^{-\lambda}} {\Gamma(1-\lambda)}
\end{equation}
When $\lambda=n$ we obtain
\begin{equation}
\label{ep3.26}
\frac {d^n U(x)} {dx^n}=\delta^{(n-1)}(x)
\end{equation}
which coincides with (\ref{ep3.7}). When $\lambda=-n$
the result is:
\begin{equation}
\label{ep3.27}
\frac {d^{-n}U(x)} {dx^{-n}}=\frac {x_+^n} {n!}
\end{equation}
In a analog way we obtain for Dirac's delta distribution:
\begin{equation}
\label{ep3.28}
\frac {d^{\lambda}\delta(x)} {dx^{\lambda}}=
\frac {x_+^{-\lambda-1}} {\Gamma(-\lambda)}
\end{equation}
\begin{equation}
\label{ep3.29}
\frac {d^n \delta(x)} {dx^n}=\delta^{(n)}(x)
\end{equation}
\begin{equation}
\label{ep3.30}
\frac {d^{-n}\delta(x)} {dx^{-n}}=\frac {x_+^{n-1}} {(n-1)!}
\end{equation}
Finally we consuder the fractional derivative of 
$e^{jbx}$:
\begin{equation}
\label{ep3.31}
\hat{f}(x)=U(x)e^{jbx}\;\;\;;\;\;\;G(s)=\frac {U[\Re(s)]} {s-ib}
\end{equation}
According to (\ref{ep2.11}):
\begin{equation}
\label{ep3.32}
\frac {d^{\lambda}U(x)e^{jbx}} {dx^{\lambda}}=\frac {1} {2\pi j}
\int\limits_{a-j\infty}^{a+j\infty}\frac {U[\Re(s)]} {s-jb}
s^{\lambda}e^{sx}ds=
\end{equation}
\begin{equation}
\label{ep3.33}
-\frac {e^{-\frac {j\pi\lambda} {2}}} {2\pi j}\int\limits_{-\infty}^{\infty}
\frac {(s+j0)^{\lambda}} {s+b+j0}
e^{-jsx} ds
\end{equation}
And thus:
\begin{equation}
\label{ep3.34}
\frac {d^{\lambda}U(x)e^{jbx}} {dx^{\lambda}}=
\frac {U(x)x^{-\lambda}} {\Gamma(1-\lambda)}
\phi(1,1-\lambda,jbx)
\end{equation}
Using (\ref{ep3.18}), (\ref{ep3.34}) transforms into:
\begin{equation}
\label{ep3.35}
\frac {d^{\lambda}U(x)e^{jbx}} {dx^{\lambda}}=
(jb)^{\lambda}U(x)e^{jbx}\left[1+\frac {\lambda} {\Gamma(1-\lambda)}
\Gamma(-\lambda,jbx)\right]
\end{equation}
When $\lambda=n$:
\begin{equation}
\label{ep3.36}
\frac {d^ne^{jbx}} {dx^n}=(jb)^nU(x)e^{jbx} 
\end{equation}
and when $\lambda=-n$:
\begin{equation}
\label{ep3.37}
\frac {d^{-n}e^{jbx}} {dx^{-n}}=(jb)^{-n}U(x)e^{jbx}
\end{equation}

\section{Circuital Application}

\setcounter{equation}{0}

As circuital application we consider a semi-infinite cable with
a voltage $V=V_0e^{j\omega t}$ applied at one end.
We use first the Fourier transform and then the Laplace 
transform for see the diferences between both treatments.

\subsection*{The Fourier Transform}

We should solve the system:
\begin{equation}
\label{ep4.1}
\begin{cases}
\frac {{\partial}^2f(x,t)} {\partial x^2}-RC\frac {\partial f(x,t)} {\partial t}=0\;\;\;;
\;\;\;x>0\\
f(0,t)=V_0 e^{j\omega t}
\end{cases}
\end{equation}
where $R$ is the resistance per unit length and $C$ is
the capacitance per unit length. Let $V(x,t)$ the voltage
along the semi-infinite cable. We use a formalism developed in
ref.\cite{tp2} to solve the system (\ref{ep4.1}). It consist in to 
define:
\begin{equation}
\label{ep4.2}
\begin{cases}
V(x,t)=U(x)f(x,t)\\
g(t)=\left.\frac {\partial f(x,t)} {\partial x}\right|_{x=0}
\end{cases}
\end{equation}
The differential equation in (\ref{ep4.1}) transforms into:
\begin{equation}
\label{ep4.3}
\frac {{\partial}^2V(x,t)} {{\partial}x^2}-RC
\frac {\partial V(x,t)} {\partial t}=
{\delta}^{'}(x)V_0e^{j\omega t}+
\delta(x)g(t)
\end{equation}
Taking the Fourier transform of (\ref{ep4.3}) we obtain:
\begin{equation}
\label{ep4.4}
\hat{V}(\alpha_1,\alpha_2)={\cal F}[V(x,t)]
\end{equation}
\[\hat{V}(\alpha_1,\alpha_2)=
\pi j V_0\delta(\alpha_1+\omega)\left[
\frac {1} {\alpha_2-\frac {1-j} {\sqrt{2}}
\sqrt{-\alpha_1 RC}}+\right.\]
\[\left.\frac {1} {\alpha_2+\frac {1-j} {\sqrt{2}}
\sqrt{-\alpha_1 RC}}\right]-
\frac {\hat{g}(\alpha_1)} {(1-j)\sqrt{-2\alpha_1 RC}}\]
\begin{equation}
\label{ep4.5}
\left[\frac {1} {\alpha_2-\frac {1-j} {\sqrt{2}}
\sqrt{-\alpha_1 RC}} -
\frac {1} {\alpha_2+\frac {1-j} {\sqrt{2}}
\sqrt{-\alpha_1 RC}}\right]
\end{equation}
Deprecating the exponential increasing in the solution
we obtain:
\begin{equation}
\label{ep4.6}
\hat{g}(\alpha_1)=-(1+j)\pi\sqrt{-2\alpha_1RC}\;\delta(\alpha_1+\omega)
\end{equation}
and then we obtain:
\begin{equation}
\label{ep4.7}
V(x,t)=V_0U(x)e^{-\sqrt{\frac {\omega RC} {2}}x}
e^{j(\omega t-\sqrt{\frac {\omega RC} {2}}x)}
\end{equation}
\begin{equation}
\label{ep4.8}
g(t)=-(1+j)\sqrt{\frac {\omega RC} {2}}\;
V_0e^{j\omega t}
\end{equation}
The current $i(x,t)$ is:
\begin{equation}
\label{ep4.9}
i(x,t)=-\frac {1} {R}\frac {\partial V(x,t)} {\partial x}\;\;\;;\;\;\;x>0
\end{equation}
As:
\begin{equation}
\label{ep4.10}
\frac {\partial V(x,t)} {\partial x}=
(1+j)\sqrt{\frac {\omega RC} {2}}
V_0e^{-\sqrt{\frac {\omega RC} {2}}x}
e^{j(\omega t-\sqrt{\frac {\omega RC} {2}}x)}
\;;\;x>0
\end{equation}
then:
\begin{equation}
\label{ep4.11}
i(x,t)=(1+j)\sqrt{\frac {\omega C} {2R}}
V_0e^{-\sqrt{\frac {\omega RC} {2}}x}
e^{j(\omega t-\sqrt{\frac {\omega RC} {2}}x)}
\;;\;x>0
\end{equation}
If we take $\lambda=1/2$ in (\ref{ep3.19} we obtain:
\begin{equation}
\label{ep4.12}
\frac {d^{\frac {1} {2}}e^{j\omega t}} {dt^{\frac {1} {2}}}=
(j\omega)^{\frac {1} {2}}e^{j\omega t}\left[1+\frac {1} {2\sqrt{\pi}}
\Gamma(-\frac {1} {2},j\omega t)\right]+
\oint\limits_{\Gamma}Z^{\frac {1} {2}}a(Z)e^{-jZt}dZ
\end{equation}
\[\frac {{\partial}^{\frac {1} {2}}V(x,t)} {\partial t^{\frac {1} {2}}}=
(j\omega)^{\frac {1} {2}}\left[1+\frac {1} {2\sqrt{\pi}}
\Gamma(-\frac {1} {2},j\omega t)\right]
e^{-\sqrt{\frac {\omega RC} {2}}x}
e^{j(\omega t-\sqrt{\frac {\omega RC} {2}}x)}+\]
\begin{equation}
\label{ep4.13}
\oint\limits_{\Gamma}Z^{\frac {1} {2}}a(Z,x)e^{-jZt}dZ
\end{equation}
Thus we have a relation between the current and the 
time derivative of the voltage:
\[i(x,t)=\sqrt{\frac {C} {R}}\left\{\left[
\frac {{\partial}^{\frac {1} {2}}} {\partial t^{\frac {1} {2}}}-
\frac {(j\omega)^{\frac {1} {2}}\Gamma(-\frac {1} {2},j\omega t)} {2\sqrt{\pi}}\right]
V(x,t)\right.-\]
\begin{equation}
\left.\label{ep4.14}
\oint\limits_{\Gamma}Z^{\frac {1} {2}}a(Z,x)e^{-jZt}dZ\right\}
\end{equation}
If we consider only the first term in the rigth side of 
(\ref{ep4.14}) we obtain the more habitual result:
\begin{equation}
\label{ep8.15}
i(x,t)=\sqrt{\frac {C} {R}}\frac {{\partial}^{\frac {1} {2}}V(x,t)}
{\partial t^{\frac {1} {2}}}
\end{equation}

\subsection*{The Laplace Transform}

If we use the Laplace transform in place of the Fourier transform
to evaluate the fractional derivatives, (\ref{ep4.12}),(\ref{ep4.13})
and (\ref{ep4.14}) are replaced by:
\begin{equation}
\label{ep4.15}
\frac {d^{\frac {1} {2}}e^{j\omega t}} {dt^{\frac {1} {2}}}=
(j\omega)^{\frac {1} {2}}e^{j\omega t}\left[1+\frac {1} {2\sqrt{\pi}}
\Gamma(-\frac {1} {2},j\omega t)\right]
\end{equation}
\begin{equation}
\label{ep4.16}
\frac {{\partial}^{\frac {1} {2}}V(x,t)} {\partial t^{\frac {1} {2}}}=
(j\omega)^{\frac {1} {2}}\left[1+\frac {1} {2\sqrt{\pi}}
\Gamma(-\frac {1} {2},j\omega t)\right]
e^{-\sqrt{\frac {\omega RC} {2}}x}
e^{j(\omega t-\sqrt{\frac {\omega RC} {2}}x)}
\end{equation}
\begin{equation}
\label{ep4.17}
i(x,t)=\sqrt{\frac {C} {R}}\left[
\frac {{\partial}^{\frac {1} {2}}} {\partial t^{\frac {1} {2}}}-
\frac {(j\omega)^{\frac {1} {2}}\Gamma(-\frac {1} {2},j\omega t)} {2\sqrt{\pi}}\right]
V(x,t)
\end{equation}
Difference between this results and the precedents is the term
that contain a contour integral.

\section{Discussion}

In this paper we have shown that Ultradistribution Theory is an 
adequate framework to define a Fractional Caculus and its 
applications. 
This definition unifies the notion of integral and derivative in one only
operation.  Several examples of application of fractional derivative
are given, including 
a circuital application: a semi-infinite cable with
a voltage $V=V_0e^{j\omega t}$ applied at one end.

\end{document}